\documentclass{aa}

\usepackage{txfonts}

\usepackage{graphicx}

\newcommand{\ergs}{ergs\,cm$^{-2}$s$^{-1}$}
\newcommand{\ergsa}{ergs\,cm$^{-2}$s$^{-1}$\AA$^{-1}$}

\newcommand{\flam}{$f_{\lambda}$}
\newcommand{\Flam}{$F_{\lambda}$}

\newcommand{\Ftio}{$F_{\rm TiO}$}
\newcommand{\fcont}{$f_{7500}$}

\newcommand{\Fcont}{$F_{7500}$}

\newcommand{\lya}{Ly$\alpha$}

\newcommand{\ic}{$I_{\rm c}$}

\newcommand{\msun}{$M_{\odot}$}

\newcommand{\mk}{$M_{\rm K}$}

\newcommand{\mh}{$[M/H]$}

\newcommand{\rc}{$R_{\rm c}$}

\newcommand{\rsun}{$R_{\odot}$}

\newcommand{\sj}{$S_{\rm J}$}
\newcommand{\sh}{$S_{\rm H}$}
\newcommand{\sk}{$S_{\rm K}$}
\newcommand{\sic}{$S_{\rm I_{\rm c}}$}
\newcommand{\src}{$S_{\rm R_{\rm c}}$}
\newcommand{\sv}{$S_{\rm V}$}
\newcommand{\Slam}{$S_{\lambda}$}
\newcommand{\slam}{$S_{\lambda}$}

\newcommand{\ten}[2]{#1\times 10^{#2}}
\newcommand{\teff}{$T_{\rm eff}$}

\newcommand{\vk}{\mbox{$V$--$K$}}

\newcommand{\irmag}{$m_{7500}-K$}

\newcommand{\yygem}{\mbox{YY\,Gem\,$\overline{\rm AB}$}}

\newcommand{\ougem}{\mbox{\object{U~Gem}}}
\newcommand{\oamher}{\mbox{\object{AM~Her}}}
\newcommand{\osscyg}{\mbox{\object{SS~Cyg}}}
\newcommand{\orupeg}{\mbox{\object{RU~Peg}}}
\newcommand{\ovcen}{\mbox{\object{V834~Cen}}}
\newcommand{\ozcha}{\mbox{\object{Z~Cha}}}

\begin{document}

\title{Barnes-Evans relations for dwarfs with an application to the determination of
distances to cataclysmic variables
\thanks{Table~2 is also available in electronic form
at the CDS via anonymous ftp to cdsarc.u-strasbg.fr (130.79.128.5)
or via http://cdsweb.u-strasbg.fr/cgi-bin/qcat?J/A+A/vol/page}
}
\author{K. Beuermann} 
%
%
\institute{ Institut f\"ur Astrophysik, Friedrich-Hund-Platz 1,
D-37073 G\"ottingen, Germany, e-mail:
beuermann@astro.physik.uni-goettingen.de }
\date{Received June 29, 2006 / Accepted August 31, 2006}
\authorrunning{K. Beuermann}
\titlerunning{Distances to cataclysmic variables}
%
\abstract { 
Barnes-Evans type relations provide an empirical relationship between
the surface brightness of stars and their color. They are widely used
for measuring the distances to stars of known radii, as the Roche-lobe
filling secondaries in cataclysmic variables (CVs).}
{ 
The calibration of the surface brightness of field dwarfs of
near-solar metalicity with spectral types A0 to L8 covers all
secondary spectral types detectable in CVs and related objects and
will aid in the measurement of their distances.}
{ 
The calibrations are based on the radii of field dwarfs measured by
the Infrared Flux Method and by interferometry. Published photometry
is used and homogenized to the Cousins \rc\ and \ic\ and
the CIT~$JHK$ photometric systems. The narrow band surface brightness
at 7500\AA\ is based on our own and published spectrophotometry. Care
is taken to select the dwarfs for near-solar metalicity, appropriate
to CVs, and to avoid errors caused by unrecognized binarity.}
{ 
Relations are provided for the surface brightness in $V$, \rc, \ic,
$J$, $H$, $K$ and in a narrow band at 7500\AA\ as functions of
\vk\ and of spectral type. The method is tested with selected CVs for
which independent information on their distances is available. 
The observed spread in the radii of early M-dwarfs of given mass or 
luminosity and its influence on the distance measurements of CVs is
discussed.}
{ 
As long as accurate trigonometric parallaxes are not routinely
available for a large number of CVs, the surface brightness method
remains a reliable means of determining distances to CVs in which a
spectral signature of the secondary star can be discerned.}

  
\maketitle

\section{Introduction} 


The spectral flux \Flam\ leaving the stellar surface is related to the
extinction corrected observed flux \flam\ at the Earth via
$F_\lambda/f_\lambda = (d/R)^2$, with $d$ and $R$ the distance and the
stellar radius, respectively.  On a magnitude scale, \Flam\ is referred to
as the surface brightness,
\begin{eqnarray}
S_{\lambda} & = & -2.5\,\mathrm{log}(F_{\lambda}/f_0) + \mathrm{const} 
\equiv M_{\lambda} + 5\,\mathrm{log}(R/R_{\odot})\\
& = & m_{\lambda} + 5\,\mathrm{log}(R/R_{\odot})-5\,{\rm log}(d/10\,{\rm pc})\\
& = & m_{\lambda} + 5\,\mathrm{log}\phi_{\mathrm mas} + 0.1564\;,
\label{slambda}
\end{eqnarray} 
with $f_0$ the flux constant at wavelength $\lambda$ for zero
magnitude, $m_\lambda = -2.5\,\mathrm{log}(f_\lambda/f_0)$ the
extinction-corrected apparent magnitude, $M_\lambda$ the absolute
magnitude, \rsun\ the solar radius, and $\phi_{\rm mas}$ the stellar
angular diameter in mas. Theoretically, \Flam\ is a function of the
effective temperature of the star with some dependence also on
metalicity and on gravity. This is the physical basis for the
empirical calibration of the surface brightness of stars with measured
angular diameters as a function of color (Barnes \& Evans 1976, Bailey
1981, Ramseyer 1994, Nordgren et al. 2002). Such a calibration allows
to determine the radius of a star of known distance and vice
versa. The dependencies on metalicity and gravity are clearly
discernible in the data (Beuermann et al. 1999) and were not properly
taken into account in the earlier calibrations of \Slam\ based on
supergiants and applied to dwarfs (Bailey 1981, Ramseyer 1994).

The surface brightness method is one of the principal avenues for
measuring distances to cataclysmic variables, binaries in which the
radius of the Roche-lobe filling dwarf secondary star is reasonably
well known from Roche geometry. Clearly, trigonometric parallaxes are
to be preferred, but are presently available only for few CVs
(Harrison et al. 1999, 2000, 2004a, McArthur et al. 1999, 2001, Monet
et al.  2003, Thorstensen 2003, Beuermann et al.  2003, 2004).  The
secondaries in CVs are characterized by a generally high level of
metalicity which we refer to loosely as `near-solar' (Beuermann
et al. 1998). Note that the abundances of individual elements may
deviate from solar.  In this paper, I present new calibrations of the
surface brightness of main sequence dwarfs with metalicities similar
to those of CV secondaries.  They supplement the relations derived for
giants, which are used in other fields of research (e.g.  Nordgren et
al.  2002). I discuss the strengths and fallacies of the
photometric/spectrophotometric methods for measuring the distances to
CVs using selected objects for which independent distance information
is available.

\section{General approach}

The establishment of a surface brightness vs. color relation requires
a set of stars of known angular radii. In this Section, I discuss
measured radii of dwarfs and compare them with the theoretical radii
of Baraffe et al. (1998, henceforth BCAH). The comparison is made in
the absolute magnitude--radius plane rather than in the mass--radius
plane since the mass does not enter the present approach. The
selection criteria of the stars used in the calibration are
subsequently explained and the stellar samples presented.

\subsection{Observed and theoretical radii of dwarfs}

Measured radii of dwarfs are available from three methods: (i) the
Infrared flux method (IRFM) (Shallis \& Blackwell 1980, Blackwell \&
Lynas-Gray, 1994, 1998, and references therein); (ii) long baseline
interferometry (Lane et al. 2001, S{\'e}gransan et al. 2003, Berger et
al. 2006); and (iii) from light curve analyses of eclipsing binaries
(Lacy 1977, Torres \& Ribas 2002, Ribas 2003, L{\'o}pez-Morales \&
Ribas 2005).

The IRFM, pioneered by Blackwell and collaborators (Shallis \&
Blackwell 1980, Blackwell \& Lynas-Gray, 1994, 1998, and references
therein), has yielded accurate radii for a large number of dwarfs and
giants. This method derives \teff, and $\phi$ from measurements of the
wavelength-integrated flux and the spectral flux in a suitable
infrared band, using model atmosphere theory in the process. Nordgren
et al. (2001) have demonstrated that the IRFM yields angular radii of
giants and supergiants which agree with interferometric radii at the
1.4\% level. A direct comparison of the IRFM radii of the dwarfs Vega,
Sirius\,A, and Atair with the interferometric radii of Mozurkewich et
al. (2003), yields a ratio of IRFM vs. interferometric radii of
$1.011\pm 0.037$. Hence, the two methods do, in fact, yield identical
results also for dwarfs. 

\begin{figure}[t]
\includegraphics[width=8.8cm]{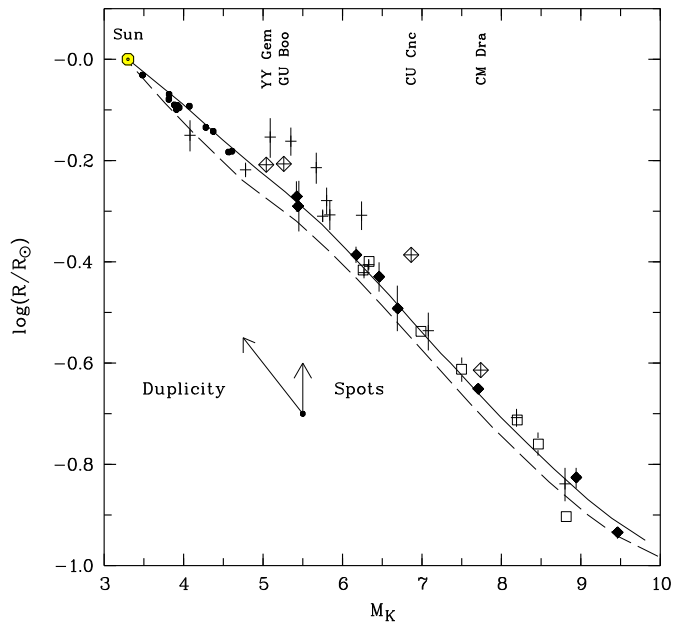}
\caption[ ]{Comparison of observed and calculated radii of dwarf
stars vs. absolute magnitude \mk\ in the $K$-band. The solid symbols
indicate radii of dwarfs with near-solar abundances (see text)
determined by the infrared flux method, open symbols are for old
disk/halo stars with metalicities [M/H] = \mbox{--0.5} to
\mbox{--2}. The Sun is indicated by the \mbox{symbol $\odot$}. Crosses
refer to interferometrically measured radii of dwarfs which have
metalicities between +0.2 and \mbox{--0.5} and in one case
\mbox{--1.0} (Gl\,191, \mk=7.08). Crossed lozenges indicate the mean
components of the four binaries named in the figure. Errors are shown
if they exceed the size of the symbols. The theoretical radii are form
Baraffe et al. (1998) for an age of 10 Gyrs and metalicities [M/H]~=~0
(solid curve) and --1 (dashed curve).}
\label{fig:radii}
\end{figure}

Figure~\ref{fig:radii} shows the radii of the Sun and twelve dwarfs
from Blackwell \& Lynas-Gray (1994, 1998) with spectral types between
G5 and K6 and metalicities between [M/H]~=~--0.50 and
+0.03\,\footnote{The metalicity [M/H] is the decadic logarithm of the
metal abundance relative to solar. I refer to metalicities
[M/H]~=~--0.4 to +0.4 as 'near solar'. The errors of the measured
metalicities are about 0.2 dex.} plotted vs. their absolute $K$-band
magnitude \mk\ (solid dots with $M_\mathrm{K}<5$). They are seen to
agree excellently with the theoretical radii of Baraffe et al. (1998)
for an age of 1--10 Gyrs and metalicity close to zero or slighly
negative.
Application of the IRFM to M-stars is more problematic because of the
increased structure in their infrared spectra and remaining
uncertainties in the theory. Leggett et al. (1996) have applied a
variant of the method to 16~M-dwarfs with well-measured parallaxes of
which eight are young disk stars with near-solar metalicities and
further eight have [M/H]~=~--0.4 to --2.0 (Fig.~\ref{fig:radii}, solid
and open lozenges, respectively). On the average, the radii of
the former exceed the BCAH radii by only 2\% (Beuermann et al. 1998,
1999), while the metal poor dwarfs seem to have radii on the average about
10\% larger than predicted for their luminosity.

The most direct method to measure stellar radii is
interferometry\footnote{Note that the derivation of limb-darkened
interferometric radii involves model atmosphere theory.}. Lane et
al. (2001), S{\'e}gransan et al. (2003), and Berger et al. (2006) have
reported radii of 14 dwarfs with spectral types K3 to M5.5 of which 13
have metalicities between \mbox{--0.46} and +0.25 and one is a
subdwarf with [M/H]~=~--1.0 (Gl\,191). Their results are added to
Fig.~\ref{fig:radii} as crosses, where the sizes of the vertical bars
indicate the \mbox{1-$\sigma$} errors. Of the 14 stars, the
interferometric radii of nine are close to the BCAH [M/H]~=~0 model
radii, while for five the interferometric radii are larger by some
20\%. These are Gl\,205, Gl\,514, Gl\,687, Gl\,752A, and Gl\,880 with
metalicities of +0.21, --0.27, +0.11, --0.05, and --0.04,
respectively, which average to solar. There are no systematic
differences between the radii determined by the different methods or
between the interferometric results of different authors. In
particular, the IRFM and interferometric radii of three dwarfs
observed with both methods agree within their errors and are also
close to the theoretical BCAH radii as exemplified in
Tab.~\ref{tab:radii}. The combined interferometric and the IRFM
results demonstrate that there is no simple relation between radius
excess and metalicity. The fact that \emph{some} stars have
substantially larger radii than suggested by the models is also
supported by the radii of dwarfs in binaries shown as crossed lozenges
in Fig.~\ref{fig:radii}. Although the effect is highly significant for
the mean components in YY~Gem (Torres \& Ribas 2002) and CM~Dra
(Metcalfe et al. 1996, Lacy 1977), the absolute differences are only
of the order of 10\% (Chabrier \& Baraffe 1995). Larger effects are
found in the binaries GU~Boo (L\'{o}pez-Morales \& Ribas 2005) and CU
Cnc (GJ\,2069Aab, Ribas 2003).

\begin{table}[b] 
\caption[]{Comparison of the radii of late-type dwarfs measured by the
Infrared Flux Method and by interferometry with the theoretical radii
of Baraffe et al. (1998) for an age of 10\,Gyr and the oberved
metalicity. The errors in the observed values are quoted below
the appropriate digits.}
\begin{tabular}{l@{\hspace{3mm}}l@{\hspace{2mm}}l@{\hspace{2mm}}cccc}
\noalign{\smallskip} \hline \noalign{\smallskip} 
Name &    \emph{SpT}  & $[M/H]\,^1$ & \mk\ & \multicolumn{3}{c}{$R/R_\odot$}\\
&&& & IRFM$\,^2$ & Interf$\,^3_.$ & BCAH$\,^4$  \\
\noalign{\smallskip}\hline\noalign{\smallskip} 
Gl\,105A& K3V  & $-0.19$       & 4.08 & 0.779          & 0.708 & 0.787 \\[-0.7ex]
        &      &\hspace{6mm}7&      & \hspace{4mm}39 & \hspace{3.5mm}  50  & \\
Gl\,411 & M1.5V& $-0.42$       & 6.33 & 0.399          & 0.393 & 0.365 \\[-0.7ex]
        &      &\hspace{6mm}7&      & \hspace{4mm}12 & \hspace{4.5mm}   8  & \\
Gl\,699 & M4V  & $-0.5$        & 8.19 & 0.194          & 0.196 & 0.176 \\[-0.7ex]
        &      &               &      & \hspace{4mm}12 & \hspace{4.5mm}   8  & \\
\noalign{\smallskip}\hline\noalign{\smallskip} 
\end{tabular}

$^1$ From Bonfils et al. (2005) and Leggett et al. (1996).\\
$^2$ From Lane et al. (2001) and S{\'e}gransan et al. (2003).\\
$^3$ From Blackwell \& Lynas-Gray (1998) and Leggett et al. (1996).\\
$^4$ From Baraffe et al. (1998), interpolated in \mk\ and \mh.
\label{tab:radii}
\end{table}

In summary, there is a sequence of stars with spectral types A0 to
M6.5 and near-solar metalicities which have radii within a few percent
of the BCAH [M/H]~=~0 model. In addition there is a spread in radius
of early M-dwarfs that is not obviously related to differences in
metalicity. Unrecognized duplicity can not be the cause for the
increased radii of some stars, because duplicity shifts the position
of a star in the $R$(\mk ) diagram approximately parallel to the
theoretical curve, at least at intermediate \mk. A plausible
explanation would be an expansion of the star due to star spots
blocking the radiative flux over part of its surface, plausible
because the stars in some of the binaries are known to possess large
spots (Torres \& Ribas 2002, L\'{o}pez-Morales \& Ribas 2005). Also,
the K-star in V471~Tau seems to have a radius $\sim18$\% larger than field K-dwarfs
due to star spots (O'Brien et al. 2001). While it is true that the
stars showing the largest radius excesses in Fig.~\ref{fig:radii},
Gl~205 (\mk=5.09), Gl~514 (\mk=5.67), Gl~752A (\mk=5.83), and Gl~880
(\mk=5.35), are not particularly active, spottedness nevertheless
seems a more likely cause of the increased radii than an abundance
effect since there is no reason for the latter to be prominent
only in early M-dwarfs. Another possible cause might be the influence
of magnetic fields on convection (Mullan \& MacDonald 2001).

A comparison between observed and theoretical radii is obviously
complicated by the spread in radii whatever the reason. Tentatively, I
consider spottedness as the cause of the increased radii of stars of a
given luminosity and effective temperature and, in what follows, I use
the term 'immaculate' for dwarfs with near-solar metalicities which
have radii near the theoretical ones (solid curve) in
Fig.~\ref{fig:radii}, irrespective of whether they are really spotless
or not. Below, I shall derive the surface brightness for immaculate
stars as a function of color or spectral type. In applying this
relation to derive distances, the possibility of increased radii of
'spotted' stars has to be specifically taken into account. For
immaculate stars of near-solar metalicity, I adopt the numerically
available theoretical magnitude-radius relation of Baraffe et
al. (1998) for an age of 10\,Gyrs and [M/H]~=~0,
$R_\mathrm{BCAH}(M_\mathrm{K})$, with a 2\% correction in the
normalization (Beuermann et al. 1998, 1999) derived from the eight
young disk M-dwarfs of Leggett et al. (1996)
\begin{equation}
R(M_\mathrm{K}) = 1.02\,R_\mathrm{BCAH}(M_\mathrm{K}).
\label{eq:radius}
\end{equation}
This relation is shown as solid curve in
Fig.~\ref{fig:radii}\,\footnote{An alternative polynomial
approximation was given by Beuermann et al. (1999) in their Eq.~(7).}
and allows to estimate the radii of immaculate dwarfs of spectral type
late M or L which presently have no measured
radii. $R_\mathrm{BCAH}(M_\mathrm{K})$ is valid down to $M_\mathrm{K}\simeq 11$
and yields a nearly constant stellar radius $R \simeq R_{\rm Jupiter}$
for still fainter objects. Its use for field L-dwarfs may not be
appropriate in individual cases and the so-derived Barnes-Evans type
relations should be considered as preliminary in the L-dwarf regime.

\subsection{Specific approach}

The Barnes-Evans relations derived here are based on the measured
radii of dwarfs with spectral types A0 to M6.5 and near-solar
metalicities and refer to immaculate (spotless) stars as discussed in
the last Section. Since the latest dwarfs with measured radii are
Gl\,551 (M5.5, interferometric) and G51-15 (M6.5, IRFM), I add dwarfs
from the list of single M-dwarfs of Henry \& McCarthy (1996) having
$V-K>6.0$ (spectral type dM5+ or later) and from the list of late M
and L-dwarfs with measured parallaxes of Dahn et al. (2002). For these
stars, the radii are estimated from Eq.~(\ref{eq:radius}) and the
angular radii follow from the well-known parallaxes. Metalicities are
taken from Cayrel de Strobel et al. (2001) or Bonfils et al. (2005) or
estimated from the position of the (single) star in the
color-magnitude diagram\footnote{A metalicity of -0.5 dex corresponds
approximately to a line 0.8\,mag below the single-star bright limit in
the $M_\mathrm{K}(V-K)$ diagram. Specifically, Gl~908, Gl~015A,
Gl~411, Gl~725A and B, Gl~643, CM~Dra~A and~B, and G3-33 are just
below this line and no longer of `near-solar' metalicity}.

\begin{figure}[t]
\vspace*{-0.6mm}
\includegraphics[width=8.72cm]{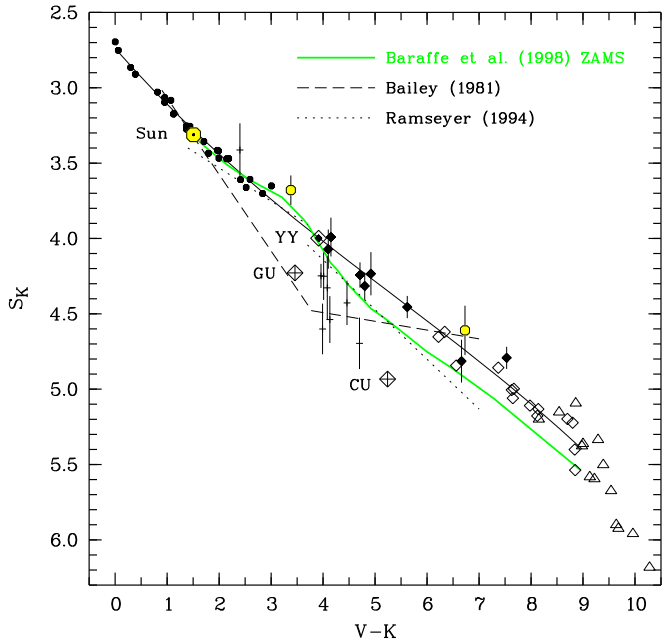}
\caption[ ]{Surface brightness \sk\ in the $K$-band vs. \vk\ for
dwarfs of near-solar metalicity. Also shown is the fit from
Eq.~(\ref{eq:skvk}) (solid line), the theoretical relation of BCAH for
stars of solar metalicity aged 10\,Gyrs (thick green/gray line), and
the relations given by Bailey (1981) (dashed) and by Ramseyer (1994)
(dotted). The subsamples included in the fit are indicated by
different symbols: dwarfs from Blackwell \& Lynas-Gray (1994,1998)
plus Vega and Sirius (solid circles); the Sun ($\odot$);
interferometric radii from Lane et al. (2001) (open circles with error
bars); YD dwarfs from Leggett et al. (1996) (filled lozenges); \yygem\
(dotted lozenge), M-dwarfs from Henry \& McCarthy (1993) (open
lozenges); M/L-dwarfs from Dahn et al. (2002) (open triangles). Shown
for comparison, but not included in the fit, are the stars with
interferometric radii (crosses) and the binaries (crossed lozenges)
from Fig.~\ref{fig:radii} with near-solar metalicity.}
\label{fig:skvk}
\end{figure}

The effect of the observed spread in the radii of early M-dwarfs on
the surface brightness is included in the Fig.~\ref{fig:skvk}, but is
omitted in the later figures and the polynomial fits are
representative of immaculate stars.

\subsection{Photometric stellar sample}

The sample of 38 single dwarfs with near-solar metalicity and
measured radii includes the following subsamples: (i) 15 dwarfs
of spectral class A0 -- K1 from Blackwell \& Lynas-Gray (1994),
supplemented with 9 K-stars from the list of ISO calibration stars
(Blackwell \& Lynas-Gray 1998) plus Vega and Sirius (Shallis \&
Blackwell 1980, Mozurkewich et al. 2003); (ii) the Sun; (iii) the mean
component of YY~Gem (Torres \& Ribas 2002); (iv) eight young disk
M-dwarfs from Leggett et al. (1996); and (v) Gl\,380 and Gl\,551 as
bona-fide immaculate stars with interferometric radii. This sample is
supplemented by (vi) 10 additional M-dwarfs with $V-K>6.0$ from the
list of single stars of Henry \& McCarthy (1993); and (vii) 36 dwarfs
of spectral type M7.5 -- L8 from the list of Dahn et al. (2002) 
with radii computed from Eq.~(\ref{eq:radius}). These supplementary
stars define the faint end of the Barnes-Evans relations. Care has
been taken to avoid unrecognized binaries. The subsample
includes both individual components of the binaries LHS2397a, Gl569B,
2M0746+20, and Kelu1, the brighter (L6) component of 2M0850+10, as
well as the mean components of DEN0205-11 and DEN1228-15. We can not
entirely exclude that a few more binaries lurk behind relatively
bright stars of the sample\footnote{T832-10443, 2M0149+29, 2M0345+25,
and 2M1439+19} or that particular youth of individual objects affects
the results. All photometric data are on the Cousins \rc, \ic\ and the
CIT~$JHK$ systems. Much of the $VR_\mathrm{c}I_\mathrm{c}JHK$
photometry has been taken from Leggett (1992), supplemented by more
recent data. Blackwell's $K$-band photometry in the Johnson system has
been converted to CIT using the transformation given by Bessel \&
Brett (1988). The total sample thus consists of 38 stars of
spectral type A0 to M6.5 which have measured radii and are referred to
as prime calibrators, supplemented by 50 stars of spectral types M5 to
L8 with radii from Eq.~(\ref{eq:radius}). Some of the L-stars have no
measured $V$-magnitude and are missing in the calibration of the
surface brightness vs. \vk. They are contained, however, in the
corresponding calibration vs. spectral type.

\begin{figure}[t]
\includegraphics[width=8.8cm]{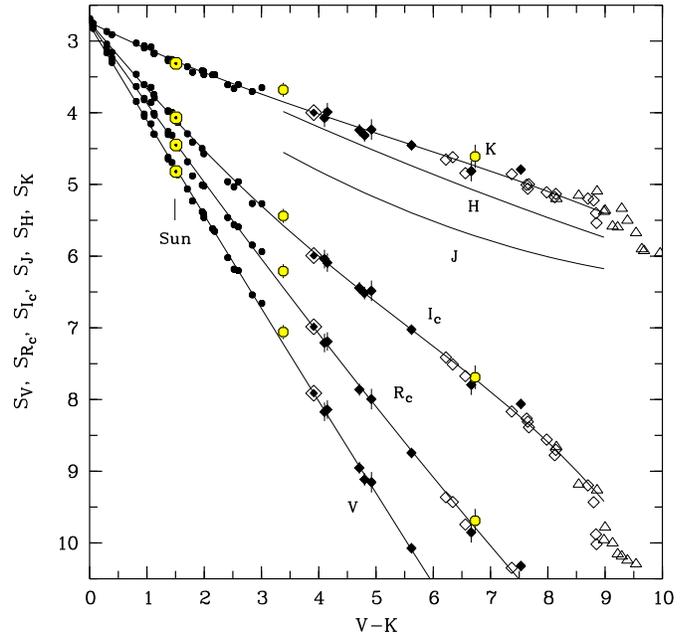}
\caption[ ]{Surface brightness \sk, \sh, sj, \sic, \src, and \sv\ for
immaculate dwarfs with near-solar metalicity. The symbols are the same
as in Fig.~\ref{fig:skvk}. The fits for the bands $K$, \ic, \rc, and
$V$ are from Tab.~\ref{tab:radii}, lines~1--4. For completeness the
fits for CIT J and H are included, but the data omitted to avoid
overlap (see text and Tab.~\ref{tab:fits}).}
\label{fig:sssvk}
\end{figure}

\begin{figure*}[t]
\begin{center}
\includegraphics[width=14.3cm]{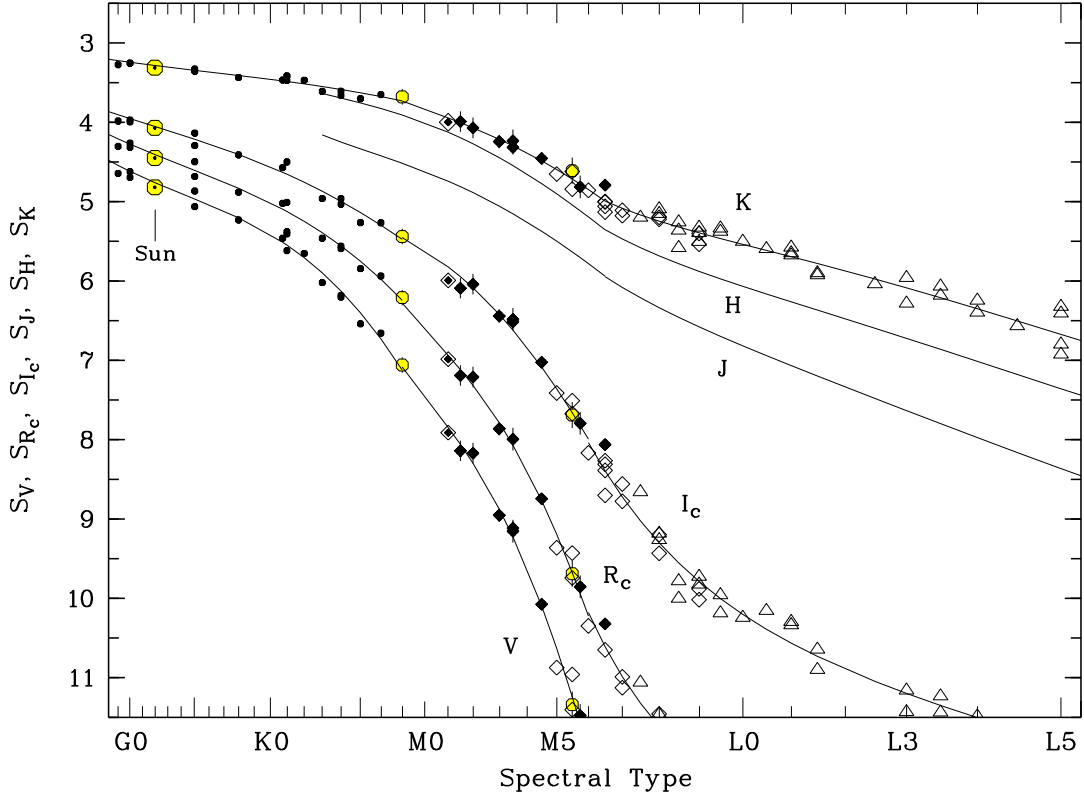}
\hfill
\raisebox{10mm}{
\begin{minipage}[b]{3.3cm}
\caption[ ]{Surface brightness \sk, \sh, \sj, \sic, \src, and \sv\ for
immaculate dwarfs with near-solar metalicity vs. spectral type. The
symbols are the same as in Fig.~\ref{fig:skvk}.  The fits are from
Tab.~\ref{tab:radii}, lines~5--15. Note that M0 follows K7. For
completeness the fits for CIT J and H are included, but the data
omitted to avoid overlap (see text and Tab.~\ref{tab:fits}). }
\end{minipage}}
\end{center}
\label{fig:sksp}
\end{figure*}

\begin{table*}[t] 
\caption[]{Parameters of least-squares fits of surface brightness
vs. \vk\ or spectral type (see text and footnote on the title page).}
\begin{tabular*}{\hsize}{rccl@{}c@{}l@{\hspace{4mm}}l@{\hspace{2mm}}l@{\hspace{2mm}}r@{\hspace{2mm}}r@{\hspace{2mm}}r@{\hspace{2mm}}r@{\hspace{3mm}}r@{\hspace{3mm}}r}
\noalign{\smallskip} \hline \noalign{\smallskip} 
(1)  & (2)  &  (3)     &\multicolumn{3}{c}{(4)}      & \multicolumn{2}{c}{\hspace{-3mm}(4)} & (5)   & (6)   & (7)   & (8)   & (9)   & (10)  \\
Line & Dep. & Indep.   &\multicolumn{3}{c}{Range of} & \multicolumn{2}{c}{\hspace{-3mm}SpT} & $a_0$ & $a_1$ & $a_2$ & $a_3$ & $a_4$ & $a_5$ \\
     & Var. & Variable &\multicolumn{3}{c}{Ind.Var.} & from &      to          &       &       &       &       &       &       \\ 
\noalign{\smallskip}\hline\noalign{\smallskip} 
1 &$S_{\rm K}$        &$V$--$K$&\hspace{1.4mm}0&--&\hspace{1.6mm}8.5 &A0&M8&   2.739 &  0.39318 &$-0.024337$&$0.00150594$             &                  &\\
2 &$S_{\rm I_{\rm c}}$&$V$--$K$&\hspace{1.4mm}0&--&\hspace{1.6mm}8.5 &A0&M8&   2.735 &  0.96065 &$-0.026441$&$-0.00455350$&$5.36349\,10^{-4}$     &\\
3 &$S_{\rm R_{\rm c}}$&$V$--$K$&\hspace{1.6mm}0  &--&\hspace{1.6mm}8.5 &A0&M8& 2.750 &  1.13296 &$-0.011563$&$-0.00020917$&                  &\\
4 &$S_{\rm V}$        &$V$--$K$&\hspace{1.4mm}0&--&\hspace{1.6mm}8.5 &A0&M8&   2.777 &  1.33603 &$-0.006106$&             &                  &\\
5 &$S_{\rm K}$        &$X$  &             58&--&21   & A0&K7         &   9.523 &$-0.68789$&  0.032091 &$-0.00076551$&$9.08518~10^{-6}$&$-4.31289~10^{-8}$\\
6 &$S_{\rm K}$        &$X$  &             21&--&13.5 & K7&M6.5       &  17.175 &$-1.73019$&  0.078260 &$-0.00125491$&                  &\\
7 &$S_{\rm K}$        &$X$  &  13.5&--&\hspace{1.6mm}2 & M6.5&L8     &   9.651 &$-0.88541$&  0.068535 &$-0.00211177$&                  &\\
8 &$S_{\rm I_{\rm c}}$&$X$  &             58&--&21   & A0&K7         &  19.534 &$-1.49122$&  0.060071 &$-0.00124465$&$1.29388~10^{-5}$&$-5.47581~10^{-8}$\\
9 &$S_{\rm I_{\rm c}}$&$X$  &             21&--&14   & K7&M6         &  36.892 &$-3.90214$&  0.164778 &$-0.00239313$&                  &\\
10 &$S_{\rm I_{\rm c}}$&$X$ &  14&--&\hspace{1.6mm}6 & M6&L4         &  14.311 &$-0.73014$&  0.061393 &$-0.00294611$&                  &\\
11 &$S_{\rm R_{\rm c}}$&$X$ &             58&--&21   & A0&K7         &  37.781 &$-3.79580$&  0.178183 &$-0.00422544$&$4.97703~10^{-5}$&$-2.33252~10^{-7}$\\
12 &$S_{\rm R_{\rm c}}$&$X$ &             21&--&14   & K7&M6         &  75.285 &$-9.65487$&  0.466505 &$-0.00777256$&                  &\\
13 &$S_{\rm R_{\rm c}}$&$X$ &  14&--&\hspace{1.6mm}6 & M6&L4         &  19.916 &$-1.83017$&  0.177753 &$-0.00690745$&                  &\\
14 &$S_{\rm V}$        &$X$ &             58&--&21   & A0&K7         &  58.160 &$-6.16978$&  0.288325 &$-0.00671549$&$7.71455~10^{-5}$&$-3.50806~10^{-7}$\\
15 &$S_{\rm V}$        &$X$ &             21&--&14   & K7&M6         &  94.817 &$-12.2938$&  0.592793 &$-0.00982453$&                  &\\
16 &$F_{7500}$        &$X$  &              48&--&37  & A0&K1         &$-284.141$& 24.7680 &$-0.656394$&  0.00649903 &                  &\\
17 &$F_{7500}$        &$X$  &             36&--&16   & K1&M4         &  60.251 &$-9.99437$&  0.513712 &$-0.00665634$&                  &\\
18 &$F_{7500}$        &$X$  &          16&--&\hspace{1.6mm}9&M4+&M9.5&$-21.130$&  6.27287 &$-0.620218$&0.02054700 &                  &\\
19 &$F_{TiO}$         &$X$  &              21&--&15  & K6&M4.5       &$-193.229$& 29.87220 &$-1.487030$&  0.02411810 &                  &\\
20 &$F_{TiO}$         &$X$  &         15&--&\hspace{1.6mm}9&M4.5&M9.5&$-1.977$ & 0.99973 &$-0.140786$&  0.00610879 &                  &\\
\noalign{\smallskip}\hline\noalign{\smallskip} 
\end{tabular*}
\label{tab:fits}
\end{table*}

\subsection{Spectrophotometric stellar sample}

The secondary stars in many CVs are detected by their TiO bands in the
red part of the spectrum. The derivation of their distances is aided
by the calibration of the narrow-band surface brightness in the quasi
continuum at 7500\AA. Specifically, I consider (i) the mean flux
between 7450 and 7550\AA\ and the flux depression below the
quasi-continuum at 7165\AA, measured by the flux difference in the
bands 7450--7550\AA\ and 7140--7190\AA\ (see Fig.~\ref{fig:ftio}a
below) corrected for atmospheric absorption. This sample is a mixed
bag of stars of which we have acquired low-resolution
spectrophotometry over the years. All spectrophotometry is readjusted
by second degree polynomials in wavelength to force agreement with the
measured photometric B, V, \rc, and \ic\ magnitudes. 

The spectrophotometric sample contains a total of 29 K6--L1 dwarfs, of
which 15 are common with the photometric sample and of these six have
measured radii. Of the remaining 14, eight are from the Henry \&
McCarthy list of single stars with $V-K<6$ and spectral types K7 to
M4+. The radii of all dwarfs without directly measured radii are from
Eq.~(\ref{eq:radius}). The complete sample contains four known
binaries (Gl65, Gl268, Gl473, Gl831) and one triple (Gl866). Their
spectral fluxes have been carefully reduced to that of the mean
or the dominant single component. One (Gl182) may be a binary
based on the brightness criterion and has been tentatively treated as
such.

For the calibration of the surface brightness at 7500\AA, the
genuine spectrophotometry of the 29 K6--L1 dwarfs has been
supplemented by synthetic 7500\,\AA\ fluxes of the Blackwell et
al. A0--K6 dwarfs with measured radii, Gl\,380 with an interferometric
radius, and the mean component of the binary YY~Gem making use of the
spectral energy distributions of dwarfs from Silva \& Cornell (1992)
for the respective spectral type adjusted to the measured \rc\
and \ic\ fluxes.

\section{Results}

\begin{figure*}[t]  
\includegraphics[width=6.8cm]{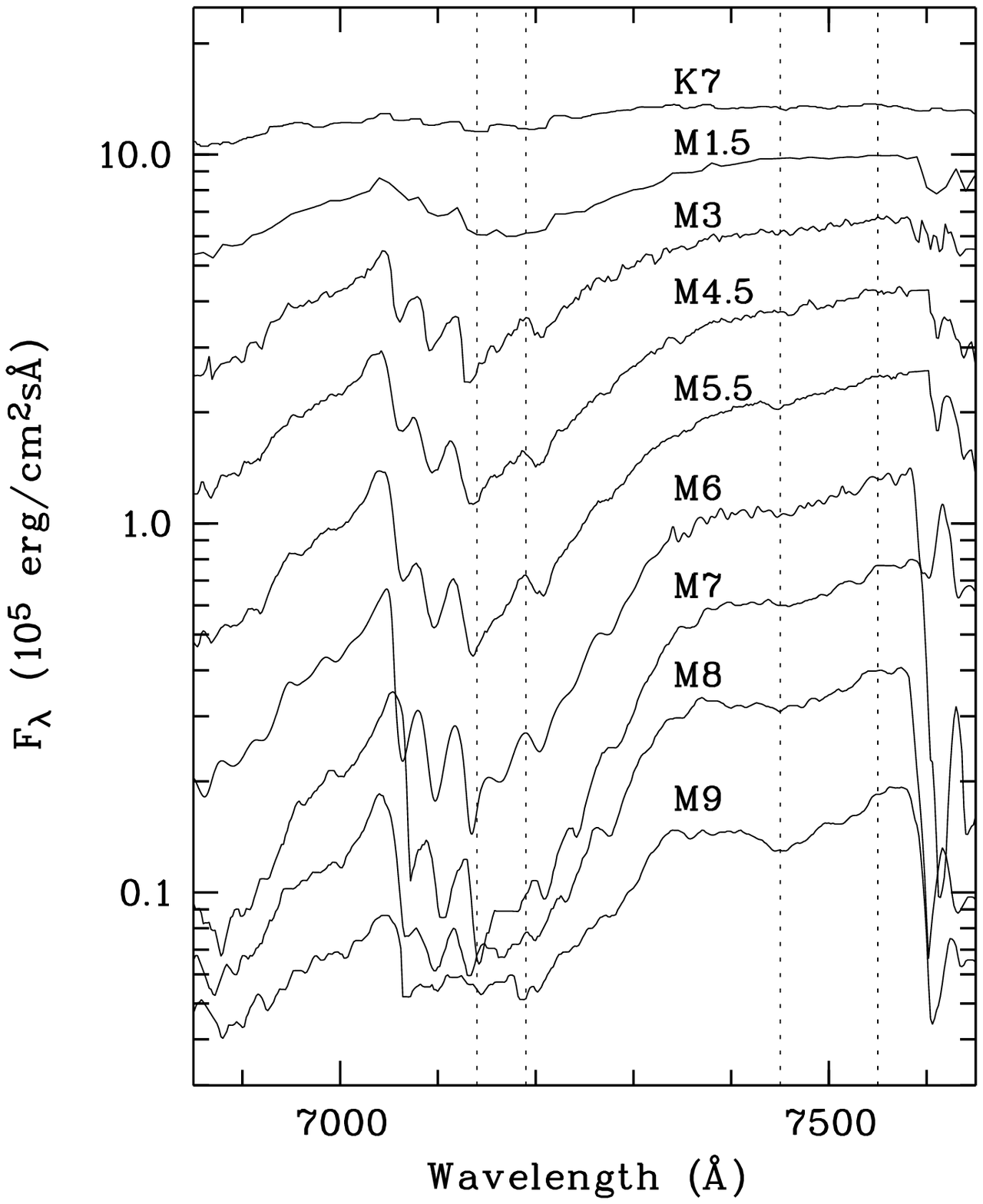}
\hfill
\includegraphics[width=10.65cm]{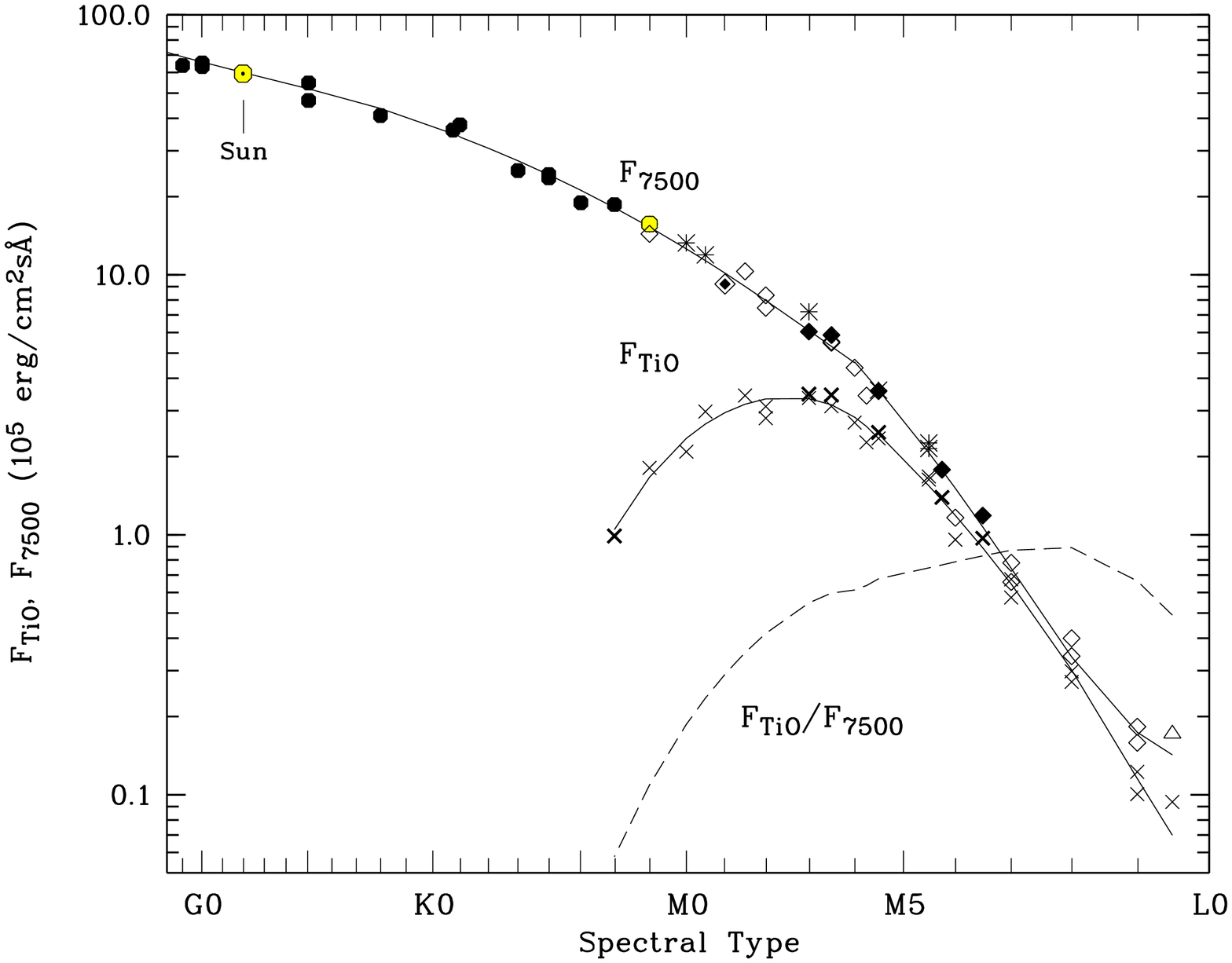}
\caption[ ]{{\it Left: } Spectral flux \Flam\ at the stellar surface
for dwarfs of spectral types K7 to M9. The vertical dotted lines
define the intervals over which the quasi-continuum at 7500\,\AA\ and
the flux deficit at 7165\,\AA, \Fcont\ and \Ftio, are averaged. {\it
Right: } Surface brightness of immaculate dwarfs with near-solar
metalicity at 7500\,\AA. Symbols are as in Fig.~\ref{fig:skvk}. The
data points include additional stars from the Henry \& McCarthy (1993)
sample of single stars (open lozenges) and the remaining stars
from our spectrophotometric sample (asterisks). Also shown is the flux
deficit \Ftio\ vs. {\it SpT} (lying crosses, lower solid curve). The
curves represent the fits from Tab.~\ref{tab:fits}, lines 16--20. }
\label{fig:ftio}
\end{figure*}

Figure~\ref{fig:skvk} shows the surface brightness \sk\ in the
$K$-band vs. \vk, with \vk\ chosen as independent variable because it
correlates well with spectral type and facilitates comparison with the
results of Bailey (1981) and Ramseyer (1994). All data with \vk$ <
8.2$ can be fitted by a third degree ($n=3$) polynomial
\begin{equation}
S_{\rm K}  =  \sum_{\mathrm{m}=0}^\mathrm{n} a_\mathrm{m}(V-K)^\mathrm{m}\\ 
\label{eq:skvk}
\end{equation}
which turns out to be nearly linear (solid black curve). The
coefficients $a_\mathrm{m}$ are given in Tab.~\ref{tab:fits},
line~1. The residuals from the fit have a standard deviation 
$\sigma=0.034$~mag for the 27 prime calibrators of spectral type
A0--K6, $\sigma=0.094$~mag for the eleven prime calibrators of
spectral types K7--M6.5, and $\sigma=0.058$~mag for all 49 stars with
$V-K<8.2$ contained in the fit. Included for comparison are the
theoretical \sk(\vk) relation of BCAH for ZAMS stars with solar
metalicity (thick green/gray curve) and the results of Bailey (1981)
and Ramseyer (1994). The present data agree almost perfectly with the
BCAH curve for $V-K<4$ demonstrating the excellent internal agreement
between the stellar radii provided by the IRFM and the theoretical
stellar radii of Baraffe et al. (1998). The theoretical curve displays
some hump structure around $V-K\simeq3$ which is also visible in the
data, but is not accounted for by the low-degree polynomial fit of
Eq.~(\ref{eq:skvk}). The divergence between the data and the BCAH
curve at larger \vk\ is a known artefact on the theoretical side
caused by the synthetic $V$-magnitudes of the late-type stellar models
coming out too bright and \vk\ correspondingly too small (I. Baraffe,
private communication). The fit of Fig.~\ref{fig:skvk} is a
substantial improvement over the widely used results of Bailey (1981)
and Ramseyer (1994) based on supergiants. The mean component of
the binary YY~Gem falls right on the fit of Eq.~(\ref{eq:skvk}), while
GU~Boo, CU~Cnc, and the dwarfs with interferometrically measured
angular diameters not included in the fit have a surface brightness
fainter by as much as 0.5~mag. That the latter reach down about to the
curve of Bailey (1981) is a coincidence. Although there is some
uncertainty in the radii attached to the L-stars, \sk\ decreases
faster than the extrapolation of the polynomial of Eq.~(\ref{eq:skvk})
at $V-K>9$ and reaches \sk~$ \sim 7.5$ for brown dwarfs with spectral
type L7--L8 and \mk\ = 12.5--12.9 (Kirkpatrick et al. 2000).

Figure~\ref{fig:sssvk} summarizes the results for the other
photometric bands. Bailey's (1981) statement that the $K$-band is best
suited for distance determinations by the surface brightness method
stays valid, because \sk\ has the shallowest slope of all photometric
bands and, hence, the smallest error in surface brightness for a given
uncertainty in \vk. In order to avoid confusion, the data for the $J$
and $H$ bands have not been included. Furthermore, since the
differences between the photometric systems are smallest in $K$, it
may be preferable to derive \sj\ and \sh\ from \sk\ using the colors
of young disk dwarfs in the appropriate color
system. Tab.~\ref{tab:colors} lists the infrared colors in the CIT
system (Leggett 1992, Stephens \& Leggett 2004). The parameters of the
polynomial fits for \sv, \src, and \sic\ are included in
Tab.~\ref{tab:fits}. They require polynomials of degree n~=~2, 3,
and 4, respectively.

Figure~4 contains the same data as a function of spectral type $SpT$,
now more densely populated in the L-star regime. This representation
is useful for applications to CVs for which the spectral type of the
secondary can be inferred from spectrophotometry, but an accurate
color \vk\ is not available. To facilitate fitting polynomials, I
express {\it SpT} by a variable $X$ as follows:
\begin{eqnarray}
{\rm A-stars:}~~SpT & = & A(58-X)\hspace{5mm}\rm{with}~~ 49 \le X \le 58, \nonumber\\ 
{\rm F-stars:}~~SpT & = & F(48-X)\hspace{5mm}\rm{with}~~ 39 \le X \le 48, \nonumber\\ 
{\rm G-stars:}~~SpT & = & G(38-X)\hspace{5mm}\rm{with}~~ 29 \le X \le 38, \nonumber\\ 
{\rm K-stars:}~~SpT & = & K(28-X)\hspace{4.5mm}\rm{with}~~ 21 \le X \le 28, \nonumber\\ 
{\rm M-stars:}~~SpT & = & M(20-X)\hspace{4mm}\rm{with}~~ 11 \le X \le 20, \nonumber\\ 
{\rm L-stars:}~~SpT & = & L(10-X)\hspace{5.5mm}\rm{with}~~~ 1 \le X \le 10.  \nonumber
\label{Xdef}
\end{eqnarray}
Note that M0 follows K7. The transformation of the variable \vk\ into
{\it SpT} (or $X$) is highly non-linear and the $S_{\lambda}$ require
piecewise representations by polynomials $S_\lambda (X) = \sum
a_\mathrm{m}X^\mathrm{m}$ of up to fith degree (Tab.~\ref{tab:fits},
lines 5--15). It is noteworthy that the scatter in \slam\ remains
practically unchanged when the spectral type via $X$ is used as the
independent variable instead of \vk, with standard deviations of
0.040~mag and 0.076~mag for the prime calibrators with spectral types
A0--K6 and K7--M6.5, respectively.

Figure~\ref{fig:ftio} (right panel) shows the results for the narrow
band surface brightness at 7500\AA\ and the flux deficiency in the TiO
band structure around at 7165\AA. The quantity plotted is the
physical flux $F_{7500}=f_{7500}(d/R)^2$ at the stellar surface, with
$f_{7500}$ the mean extinction-corrected observed flux in the
7450--7550\AA\ band in \ergsa. Correspondingly, $F_\mathrm{Tio}$ is
the difference between the mean surface fluxes in the bands
7450--7550\,\AA\ and 7140--7190\,\AA, as depicted in the left panel of
Fig.~\ref{fig:ftio}.  \Fcont\ decreases continually for dwarfs of
spectral type $A$ to $L$, while \Ftio\ assumes a maximum at M2--M3 and
vanishes for stars earlier than K6. The stars used for \Ftio\
are a subset of those for \Fcont, with the fat crosses referring to
stars with measured radii and the thin crosses to stars with radii
from Eq.~(\ref{eq:radius}). The ratio \Ftio/\Fcont\ reaches a maximum
at spectral type dM7 and tends to decrease again for still later
spectral types because the flux at 7500\AA\ is increasingly depressed
by VO absorption (dotted curve). Tab.~\ref{tab:fits}, lines 16--20
provide the coefficients of piecewise cubic polynomial fits,
$F_\mathrm{7500,TiO} = a_0 + a_1\,X + a_2\,X^2+ a_3\,X^3$, as
functions of~$X$ in units of $10^5\,\mathrm{erg\,cm}^{-2}{\mathrm
s}^{-1}\AA^{-1}$.

\begin{table}[t] 
\caption[]{Intrinsic infrared colors of young disk dwarfs (CIT system).}
\begin{tabular}{ccccccccc}
\noalign{\smallskip} \hline \noalign{\smallskip} 
{\it SpT} & \vk\ &$J$--$K$&$H$--$K$&&{\it SpT}&\vk\ &$J$--$K$&$H$--$K$ \\
\noalign{\smallskip}\hline\noalign{\smallskip} 
 M0 & 3.47 & 0.85 & 0.17 & &  M7 & 7.85 & 0.96 & 0.37 \\ 
 M1 & 3.76 & 0.85 & 0.18 & &  M8 & \hspace{-1mm}8.7  & 1.05 & 0.43 \\ 
 M2 & 4.05 & 0.86 & 0.20 & &  M9 & \hspace{-1mm}9.0  & 1.12 & 0.47 \\ 
 M3 & 4.60 & 0.85 & 0.23 & &  L1 & \hspace{-1mm}9.5  & 1.25 & 0.51 \\ 
 M4 & 5.18 & 0.87 & 0.26 & &  L3 & \hspace{-3mm}10.0 & 1.51 & 0.66 \\ 
 M5 & 5.79 & 0.87 & 0.30 & &  L5 & \hspace{-3mm}11.5 & 1.72 & 0.74 \\ 
 M6 & 6.65 & 0.89 & 0.33 & &  L8 &      & 1.72 & 0.74 \\                 
\noalign{\smallskip}\hline\noalign{\smallskip} 
\end{tabular}
\label{tab:colors}
\end{table}

As a further utility, I define a color 
\begin{equation}
m_{7500} - K = -2.5\,{\rm log}\,f_{7500} - K - 22.15
\end{equation}
which measures the ratio between \fcont\ and the $K$-band flux
(Tab.~\ref{tab:useful}).  As usual, the flux constant is chosen such that
\mbox{$m_{7500} - K$} = 0.0 for spectral type A0.  This color allows a
quick estimate of the $K$-band magnitude of a dwarf detected by its
7500\AA\ flux and vice versa. 

The results presented in this Section are strictly applicable only to
dwarfs. For objects related to CVs, as supersoft X-ray sources and
low-mass X-ray binaries including black hole binaries, the calibration
of \sk(\vk) can still be used if the secondary is a Roche-lobe filling
subgiant or even a giant. The calibrations of the surface brightness
in the $V, R_\mathrm{c}$, and $I_\mathrm{c}$ bands as functions of color
deviate increasingly from those of dwarfs for decreasing gravity and
do so even more if considered as a function of spectral type.

\begin{table}[t] 
\caption[]{Color \irmag\ characterizing the spectral energy distribution
of main-sequence stars of luminosity class V (see text).}
\begin{tabular}{lcllcllc}
\noalign{\smallskip} \hline \noalign{\smallskip} 
{\it SpT} & \irmag & & {\it SpT} & \irmag  & & {\it SpT} & \irmag\\
\noalign{\smallskip}\hline\noalign{\smallskip} 
A0  &  0.00 & & M0  &  1.97 & & M6  &  3.34 \\
F0  &  0.42 & & M1  &  2.10 & & M7  &  3.77 \\
G0  &  0.79 & & M2  &  2.23 & & M8  &  4.38 \\
K0  &  1.20 & & M3  &  2.39 &  & M9  &  5.00 \\	  
K3  &  1.42 & & M4  &  2.61 &  & L0  &  \hspace{1mm}5.46:\\
K7  &  1.88 & & M5  &  2.94 &  & L1  &  \hspace{1mm}5.65:\\
\noalign{\smallskip}\hline\noalign{\smallskip} 
\end{tabular}
\label{tab:useful}
\end{table}

\section{Distances to CVs}

In what follows, I compare the distances of a small sample of five CVs
obtained by different methods including trigonometry. The flux
observed from the Roche-lobe filling secondary star in a CV depends on
aspect and, hence, the radius appearing in Eq.~(2) is $R_\mathrm{eff}
= (A/\pi)^{1/2}$, with $A$ the cross section presented by the
secondary star at the orbital phase of the observation. The secondary
is best observed in a low state of switched-off accretion when other
light sources and the illumination of the secondary are minimal. Even
then, in the absence of heating effects, the spectral type varies
slightly with aspect as a result of von Zeipel's law (e.g., Reinsch et
al. 2006). The equivalent volume radius of the Roche lobe filling
secondary is
\begin{equation} 
R_2/R_{\odot} = 0.234\,(M_2/M_{\odot})^{1/3}\,P^{2/3}f(q)
\label{radius}
\end{equation} 
\noindent where $M_2$ is the mass of the secondary star, $P$ is the
orbital period in hours, and $f(q)$ varies between 0.980 and 1.031 for
$q \le 1.0$ (Kopal 1959). For a given inclination $i$ of the system
and a given orbital phase~\footnote{Phase $\varphi=0$
refers to inferior conjunction of the seondary star.} $\varphi$, the ratio
$R_\mathrm{eff}/R_2$ can be computed from the dimensions of the
Roche-lobe filling star tabulated by Kopal or calculated numerically
from a Roche lobe model.
The distance $d$ of a CV is obtained from Eq.~(2) with $R = R_\mathrm{eff}$ as
\begin{equation} 
\mathrm{log}\,d = (m_{\lambda} - A_{\lambda}- S_{\lambda})/5 + 1 +
\mathrm{log}(R_\mathrm{eff}/R_{\odot}),
\label{distance}
\end{equation} 
with $m_{\lambda}$ the observed magnitude of the secondary and
$A_{\lambda}$ the extinction. The error budget of $d$ is based on an
assumed systematic uncertainty of $\pm 10$\% in the calibration of the
surface brightness added quadratically to the systematic and
statistical uncertainties in the measured extinction-corrected flux of
the secondary star taken to be 10\% or 0.10\,mag if not otherwise
quoted in the text. On top of this, $d$ varies as $M_2^{1/3}$. Results
are summarized in Tab.~\ref{tab:distances}. The quoted photometric
distances carry the band on which they are based as a subscipt, e.g.,
$d_{\mathrm V}$, $d_{\mathrm K}$, or $d_{7500}$ for the distances
deduced from $V$, $K$ and \fcont, respectively.

\begin{table*}[t] 
\caption[]{Examples of distances of CVs determined by the surface
brightness method.  Columns (1) to (4) contain information of the
secondary star, column (5) the contribution of the secondary star to
the observed flux at 7500\AA, columns (6) and (7) the $V$ and
$K$-magnitudes of the secondary star, column (8) the mass used in
calculating the distances $d_\mathrm{7500}$, $d_{\rm V}$, and $d_{\rm
K}$ in columns (9), (10), and (11), and column (12) gives the
trigonometric distance if available. See text for errors of the
quantities in columns (4) to (8).}
\begin{tabular*}{\hsize}{@{\extracolsep{\fill}}llclccccccccl}
\noalign{\smallskip} \hline \noalign{\smallskip} 
(1) & (2) & (3) & (4) & (5) & (6) & (7) & (8) & (9) & (10)& (11) & (12) \\
Name & Type & $P_{\mathrm orb}$ & Sp.T. & $f_{\mathrm 7500}$ & $V$ & $K$ & $M_2$ &
$d_\mathrm{7500}$ & $d_{\mathrm V}$ & $d_{\mathrm K}$ & $1/\pi$ \\
 && (h) &  &(\ergsa) & (mag)& (mag)  & $(M_{\odot})$ &
(pc) & (pc)& (pc) & (pc)\\
\noalign{\smallskip}\hline\noalign{\smallskip} 
\multicolumn{11}{l}{\sl (1) \qquad CVs with trigonometric distances:}\\[0.5ex]
U Gem&DN&4.246&M4+&$\ten{4.6}{-15}$ &       &$10.95$&0.41&$97\pm7$&            &$94\pm 7$&\hspace{-1.5mm}$100\pm4$\\
AM Her&AM&3.094&M4-&$\ten{2.7}{-15}$&       &$11.79$&0.20&$88\pm8$&            &$89\pm 8$&\hspace{2mm}$83\pm5^{\,1)}$\\
SS Cyg&DN&6.603&K4&    &\hspace{-2mm}$12.7 $&       &0.80&        &$156\pm17$  &         &$165\pm12$\\
RU Peg&DN&8.990&K3&                 &$13.35$&       &0.94&        &$299\pm 50$ &         &$299\pm24$\\[0.5ex]
\multicolumn{10}{l}{\sl (2) \qquad CVs with \fcont\ and $K$-band measurements:}\\[0.5ex]
V834Cen&AM&1.692&M$5.5$&$\ten{2.3}{-16}$&&$13.85$&0.110&\hspace{0mm}$110\pm7$&&\hspace{0mm}$104\pm6$&\\
Z Cha&SU&1.788&M6&$\ten{1.6}{-16}$  &&$14.03$&0.125&$114\pm8$&&$112\pm8$&\\
\noalign{\smallskip}\hline\noalign{\smallskip}
\end{tabular*}
\footnotesize{$^1)$ Mean of the trigonometric distances by Thorstensen
(2003) and C.~Dahn, $\pi=85\pm 5$\,pc, private communication.}
\label{tab:distances}
\end{table*}

\subsection{CVs with trigonometric distances}

\subsubsection{U Geminorum}

\ougem\ has a well determined HST parallax of $100\pm4$\,pc (Harrison et
al. 1999, 2001, 2004a). The secondary star is of spectral type M4+
(Stauffer et al. 1979, Wade 1979, Friend et al. 1990) and is
prominently seen against the rather faint quiescent accretion
disk. Wade's (1979) spectrophotometry and his assumption of a
frequency-independent spectral flux $f_{\nu}$ from the accretion disk
yields \fcont\,= $\ten{4.6}{-15}$\,\ergs\ for the secondary star at
$\varphi = 0.08$. Panek \& Eaton (1982) find a minimum brightness at
$\varphi=0$ of $H=11.35$ and $K=11.04=K_\mathrm{CIT}$, which also
accounts for the eclipse of at least the bright central part of the
accretion disk. These quantities imply $m_{7500} - K = 2.65$ as
expected for the spectral classification. At $i=69^\circ$ and
$\varphi=0$, $R_\mathrm{eff}=0.98\,R_2$. I adopt $M_2 =
(0.41\pm0.02)\,$\msun\ (Long \& Gilliland 1999) which equals the mass
of a Roche-lobe filling main sequence star (Patterson 1984). The
distances derived from \fcont\ and $K$ are $d_\mathrm{7500}=(97\pm 7)$\,pc and 
$d_\mathrm{K}=(94\pm 7)$\,pc in good agreement with the HST value.
Taken at face value, this agreement suggests that the appearence of
the secondary star in \ougem\ is not strongly affected by spottedness.

\subsubsection{AM Herculis} 

Trigonometry has yielded distances of \mbox{$79^{+8}_{-6}$\,pc}
(Thorstensen 2003) and \mbox{$(85\pm 5)$\,pc} (USNO parallax, C. Dahn,
private communication). G\"ansicke et al. (1995, see their Fig.~6)
discussed the blue and red spectrophotometry of Schmidt et al. (1981)
and the available low state photometry of \oamher. The adjusted
spectra of Gl273 (dM3.5) and G3-33 (dM4.5) fit the TiO bands in the
visible and indicate a visual magnitude of the secondary of
$V$\,=\,$16.8$. The 7500\AA\ spectral flux $f_{7500}=
\ten{2.7}{-15}$\,\ergsa\ (uncertainty $\pm 15$\%) indicates a spectral
type between those of the two comparison stars. Low state infrared
photometry (Bailey et al. 1988) yields a $K$-magnitude of the
secondary converted to the CIT system of 11.79. Furthermore, low-state
spectrophotometry (Bailey et al. 1991) yields $K=11.60$ close to
$\varphi=0$ which is almost entirely from the secondary. The implied
\vk\ color yields a spectral type dM$4-$ to dM4
(Tab.~\ref{tab:colors}). The mass of the secondary is not well
constrained. The mass ratio $q = M_2/M_1 = (0.47\pm0.05)$ (Southwell et
al. 1995) and the primary mass $M_1 = (0.65\pm 0.16)$\,\msun\
(G\"ansicke et al. 2006) suggest a lower limit $M_2\simeq0.20$\,\msun,
close to the secondary mass at which CVs get in contact again after
crossing the period gap. Such a low mass is not unreasonable given the
facts that (i) \oamher\ seems to be at the brink of entering the gap and
(ii) its secondary has a later spectral type than a Roche-lobe filling
main sequence secondary suggesting that some bloating has taken place
(Beuermann et al. 1998, their Fig.~5). The latter argument
implies a mass below the main sequence mass in a CV with an orbital
period of 3.1\,h, $M_2=0.28$\,\msun\ (Patterson 1984). Given the
intermediate inclination of $35^\circ-50^\circ$ (G\"ansicke et
al. 2001, and references therein), I use $R_\mathrm{eff} \simeq
R_2$. Assuming an M4- secondary and neglecting extinction, I obtain
\mbox{$d_{7500} = (88\pm8)(M_2/0.20\,\mathrm{M_\odot})^{1/3}$\,pc} and
$d_\mathrm{K} = (89\pm 8)(M_2/0.20\,\mathrm{M_\odot})^{1/3}$\,pc, 
just compatible with the trigonometric distance for $M_2$ near
0.2\,\msun. This derivation, however, assumes that the secondary is
close to immaculate, while Hessman et al. (2000, see their Fig.~6)
suggested it to be heavily spotted. The two distance estimates
decrease to the trigonometric value for a surface brightness fainter
by a moderate 0.2~mag due to spottedness.

\subsubsection{SS Cygni}

The HST distance of \osscyg\ (Harrison et al. 1999, 2000, 2004a) is $(165\pm
12)$\,pc while Bailey (1981) quoted a $K$-band distance of only 87\,pc.
SS~Cyg has $K=9.4$ in quiescence which Bailey (1981) assumed to be
entirely due to the secondary star. The accretion disk, however,
contributes substantially to the total flux (Harrison et
al. 2000). Wade (1982) synthesized the observed flux distribution and
quoted $V=12.7$ for the secondary at the total quiescent flux level of
$V=11.7$. The stellar components are both rather massive and
$M_2 \simeq 0.8$\,\msun. With a median spectral classification of the
secondary of K4 (Beuermann et al. 1998) and $R_\mathrm{eff}=1.01\,R_2$
at $i\simeq 40^\circ$, the calibration of the visual surface
brightness yields $d_\mathrm{V}=(156\pm 17)(M_2/0.8)^{1/3}$\,pc,
roughly consistent with the HST result.  Since $d_\mathrm{V}$ is
nominally smaller than the trigonometric distance, a reduced surface
brightness of a potentially spotted secondary provides no remedy.

Wade's (1982) flux synthesis and the spectral type K4 imply that the
secondary has $K\simeq 10.1$ and contributes only 50\% to the observed
infrared flux. This case demonstrates the pitfalls of using an
insufficiently secured $K$-band magnitude of the secondary star for
distance measurements. Given the concerns expressed with respect to a
distance as large as 165\,pc (Schreiber \& G\"ansicke 2002), it might
be useful to re-evaluate the contribution of the secondary to the
observed absolute spectral energy distribution and to obtain more
accurate masses of both stellar components.

\subsubsection{RU Pegasi}

\orupeg\ is another bright long-period CV with an accurate HST parallax
of $\pi_\mathrm{abs}=3.35\pm 0.26$\,mas (Harrison et
al. 2004b). Again, the accretion disk contributes to the observed
visual and infrared flux. The spectral type of the secondary star is
K2 or K3 (Wade 1982, Friend et al. 1990, Harrison et
al. 2004b). Extinction seems to be negligible. \orupeg\ has a mean
visual magnitude in quiescence of $V=12.62$ with a light curve
dominated by flickering (Bruch \& Engel 1994, Bruch, private
communication, 2006) and going down to $V=13.1$ (Stover 1981). Hence
the secondary star can not contribute more than 63\% of the mean
visual light level. Wade's (1982) flux synthesis allows for a
contribution to the observed visual flux of 61--90\% by a K2 star and
38--75\% by a K3 star (see Wade's Table III), favouring the spectral
type K3V. I adopt a contribution of $(50\pm 12)$\% corresponding to a
visual magnitude of the secondary star of $V=13.35\pm 0.25$. In the
infrared, \orupeg\ has $K=10.48$ (Harrison et al. 2004b) which can
tentatively be synthesized from $K_\mathrm{sec}=10.93$ and
$K_\mathrm{disk}=11.65$ for the contributions by the secondary star
and disk (plus any other light source). Here, I have used $V-K=2.42$
for a K3 dwarf. There is some controversy about the mass of the
secondary star which results from the ill-known inclination. Ritter \&
Kolb (2003) quote $M_2=(0.94\pm 0.04)$\,\msun\ from Shafter's (1983)
thesis, while Friend et al. (1990) derive $(1.07\pm0.02)$\,\msun. The
Roche lobe radii for the two masses are 1.01 and 1.06\,\rsun,
respectively, suggesting that the secondary is right on the main
sequence for the larger and minimally evolved for the smaller
mass. The visual surface brightness of a spotless K3 dwarf is
$S_\mathrm{V}=6.0$. The distance from the visual magnitude then is
$d_\mathrm{V}=(299\pm 50)$\,pc for $M_2=0.94$\,\msun\ and
$d_\mathrm{V}=(313\pm 50)$\,pc for $M_2=1.06$\,\msun. The quoted
magnitude $K_\mathrm{ses}$ is not an independent quantity and yields
the same distance. The result is in excellent agreement with the
trigonometric parallax and the error of 50\,pc from the combined
uncertainties in the visual flux and the mass of the secondary is
twice that of the parallax error. 

\subsection{Other CVs}

The two short-period CVs considered in this Section have no
trigonometric distances, but independent distances can be estimated
from the spectral fluxes at 7500\,\AA\ and in the $K$-band.

{\it V834 Centauri:~} Beuermann et al. (1989) and Puchnarewicz et
al. (1990) obtained spectrophotometry of \ovcen\ in the low state and
found $SpT \simeq \,$dM5 and dM6, respectively. The 7500\AA\ flux of
the secondary is \fcont\,= $2.2\,10^{-16}$\,\ergs. Since some
cyclotron emission may still be present in the low state, I adopt the
minimal infrared flux to represent the secondary, $K=(13.85\pm0.08)$
(Sambruna et al. 1991). The color \irmag\ agrees with that expected
for a spectral type dM5.5. \mbox{I use} $i \simeq 50^{\circ}$ as
suggested by the light curve, $R_\mathrm{eff}\simeq R_2$ at
$\varphi=0$, and assume the secondary to be an immaculate main
sequence star with $M_2 = 0.11$\,\msun\ (Patterson 1998, his
Eq.~5). Both approaches yield the nearly the same distance,
$d_{7500}=(110\pm 7)$\,pc and $d_{7500}=(104\pm 7)$\,pc, which scale as
$(M_2/0.11M_\odot)^{1/3}$.

{\it Z Chamaeleontis:~} Bailey et al. (1981) performed phase-resolved
infrared photometry of the dwarf nova \ozcha\ in quiescence and Wade \&
Horne (1988) obtained spectrophotometry of the dM5.5 secondary star in
eclipse. The latter is still contaminated by emission from the
uneclipsed outer accretion disk and the same may hold for the infrared
fluxes in eclipse. The dM5.5 spectrum adjusted by Wade \& Horne to the
eclipse spectrum has \fcont\,$=\ten{1.6}{-16}$\,\ergs (0.3 mJy), and
the CIT $K$-band magnitude in eclipse is 14.03. The resulting
\irmag=3.30 is suggestive of a spectral type M6. Alternatively, if
M5.5 is correct, either $K=14.20$ or
\fcont\,$=\ten{1.85}{-16}$\,\ergs. The latter would imply that the TiO
features in the secondary are slightly weaker than those in the
comparison star Proxima Cen (Gl~551) also on the unilluminated
hemisphere, the former that some infrared disk flux is still seen at
eclipse center. Both possibilities can not be excluded. Wade \& Horne
determined a $M_2=(0.125\pm0.014)$\,\msun\ which implies a radius
$R_2=(0.171\pm0.006)$\,\rsun. With $i = 82^\circ$ and
$R_\mathrm{eff}=0.96\,R_2$ at $\varphi=0$, the different flux
combinations yield distances $d_{7500}\simeq d_\mathrm{K}$ between 111
and 130 pc for $M_2=0.125M_\odot$.  Tab.~\ref{tab:distances} quotes
the distances resulting from the nominal fluxes using the surface
brightness of an immaculate M6 dwarf. They scale as
$(M_2/0.125M_\odot)^{1/3}$.

\section{Discussion and conclusion} 

I have presented new calibrations of the surface brightness of field
dwarfs of spectral types A to L with near-solar metalicity. The
extension to late L-dwarfs is important considering the
increasing evidence that the secondaries in short-period CVs are
substellar. These Barnes-Evans relations presented here refer to
immaculate (spotless) stars which show a fair agreement with
theoretical radii for near-solar metalicity (Baraffe et al. 1998,
S{\'e}gransan et al. 2003).
A caveat to be kept in mind in using the surface brightness
method is the spread in the radii of field stars of spectral type early
M and a given absolute magnitude found from interferometric
observations (S{\'e}gransan et al. 2003, Berger et al. 2005) and light
curve analyses of eclipsing binaries (Metcalfe et al. 1996, Torres \&
Ribas 2002, Ribas 2003, L{\'o}pez-Morales \& Ribas 2005). The ultimate
cause for this spread in radius is still debated, but spottedness is a
distinct possibility.

The comparison of the different methods to determine distances to CVs
shows a satisfactory internal consistency, provided the contributions
by other light sources in the systems are estimated realistically or
are minimized by observing the systems in low states of switched-off
accretion. None of the examples in the present limited study suggests
a significant deviation of the properties of the secondary stars from
the immaculate variety, since the distances obtained assuming such
stars agree within errors with the trigonometric distances. Only in
the case of the M4V secondary star in \oamher\ may the actual surface
brightness fall some 0.2 mag below that of an immaculate M4 dwarf, not
an unreasonable result considering that the secondary may be heavily
spotted (Hessman et al. 2000). The spread in the radius of field stars
translates into a spread in the surface brightness of the Roche-lobe
filling CV secondaries of a given spectral type or color. For a
heavily spotted star, the surface brightness may fall a couple of
tenths of a magnitude below the values from the polynomial fits
reported here.

Another systematic difference between field stars and CV secondaries 
might be age. The latter are typically older than 1~Gyr and, with
perhaps a few exceptions, that seems to hold also for all M/L-dwarfs
in our sample suggesting that systematic age differences in the
populations do not prevail.

A more global view of the problem of CV distances should include the
white dwarf component which, in some CVs, is more easily identified in
the far UV than the secondary star in the infrared. In principle, fits
of stellar atmosphere spectral models to the \lya\ profile can yield
the temperature and the radius of the white dwarf (via log\,$g$) and,
therefore, an independent distance measurement (Araujo-Betancor et
al. 2005). This is a variant of the surface brightness method.

As long as trigonometric parallaxes do not become routinely available
also for the fainter CVs, a coordinated effort involving all available
methods to measure distances can lead to improved space densities and
to a better understanding of the evolution of CVs.

\acknowledgements{I thank the anonymous referee for helpful
comments which led to an improved presentation of the results. 
M. Weichhold contributed to the early stages of this work. I
benefitted from many discussions with Isabelle Baraffe, Ulrich Kolb,
Boris G\"ansicke, Frederick Hessman, Klaus Reinsch, and Axel
Schwope. Sandra Leggett, Klaus Reinsch and Axel Schwope generously
provided part of the spectrophotometric data used in the
calibrations.}


\end{document}